\documentclass[10pt,conference]{IEEEtran}
\usepackage[T1]{fontenc}
\usepackage{cite}
\usepackage{xcolor}
\usepackage{amsmath,amssymb,amsfonts}
\usepackage{algorithmic}
\usepackage{graphicx}
\usepackage{textcomp}
\usepackage{float} 
\usepackage{subfigure}
\usepackage{array}
\usepackage{booktabs}
\usepackage[ruled,linesnumbered]{algorithm2e}
\usepackage{verbatim}

\def\BibTeX{{\rm B\kern-.05em{\sc i\kern-.025em b}\kern-.08em
		T\kern-.1667em\lower.7ex\hbox{E}\kern-.125emX}}

\begin{document}
	
	\title{Edge-Assisted Lightweight Region-of-Interest Extraction and Transmission for Vehicle Perception }
	\author{Yan Cheng, Peng Yang,~\IEEEmembership{Member,~IEEE},~Ning Zhang,~\IEEEmembership{Senior~Member,~IEEE},~Jiawei Hou}
	
	\author{\IEEEauthorblockN{Yan~Cheng\IEEEauthorrefmark{1},~Peng~Yang\IEEEauthorrefmark{1},~Ning Zhang\IEEEauthorrefmark{2}, and~Jiawei Hou\IEEEauthorrefmark{1}}
		\IEEEauthorblockA{\IEEEauthorrefmark{1}School of Electronic Information and Communications, Huazhong University of Science and Technology, Wuhan, China \\
			\IEEEauthorrefmark{2}Department of Electrical and Computer Engineering, University of Windsor, Windsor, ON, Canada \\}
		Email: \IEEEauthorrefmark{1}\{y$\_$cheng, yangpeng, jerry$\_$hou\}@hust.edu.cn, \IEEEauthorrefmark{2}ning.zhang@uwindsor.ca}
	
	\maketitle
	\begin{abstract}
		
		To enhance on-road environmental perception for autonomous driving, accurate and real-time analytics on high-resolution video frames generated from on-board cameras becomes crucial. In this paper, we design a lightweight object location method based on class activation mapping (CAM) to rapidly capture the region of interest (RoI) boxes that contain driving safety related objects from on-board cameras, which can not only improve the inference accuracy of vision tasks, but also reduce the amount of transmitted data. Considering the limited on-board computation resources, the RoI boxes extracted from the raw image are offloaded to the edge for further processing. Considering both the dynamics of vehicle-to-edge communications and the limited edge resources, we propose an adaptive RoI box offloading algorithm to ensure prompt and accurate inference by adjusting the down-sampling rate of each box. Extensive experimental results on four high-resolution video streams demonstrate that our approach can effectively improve the overall accuracy by up to 16\% and reduce the transmission demand by up to 49\%, compared with other benchmarks.
	\end{abstract}
	
	\section{Introduction}
	The potential of autonomous driving in reducing traffic congestion and improving driving safety has attracted much attention. Its core function components are outlined as sensing, planning, and control \cite{AutonumousDriving}. Highly accurate environmental sensing is fundamental to decision making in autonomous driving, especially in complex road segments and hazardous weather conditions \cite{Ale2019Delay}. However, limited by the sensing range and accuracy of on-board sensors, most off-the-shelf vehicles can only reach the level of L2 in autonomous driving \cite{Survey, yang23edge}.
	
	High-resolution video cameras can help to capture rich on-road information and improve the inference accuracy of perception tasks based on high-performance compute \cite{Remix, wang22object}. Besides, advanced convolutional neural networks (CNNs) can effectively process video streams compared to traditional feature extraction methods, but require significant amount of accelerators, such as central/graphics processing units (CPUs/GPUs). Most existing CNNs have been developed to deal with low-resolution images to increase processing speed, such coarse-grained down-sampling of high-resolution input can significantly reduce the recognition accuracy \cite{HOU, zhou2023bandwidth}. Consequently, taking full advantage of the rich information of high-resolution videos while meeting real-time requirements is one of the challenges in video analytics. Moreover, self-driving vehicles are sensitive to latency, as it is directly related to driving safety. The massive amounts of data generated by multiple high-resolution cameras deployed inside and outside the vehicle pose tremendous pressure on real-time video analytics system \cite{ELF, EdgeDuet, yang20edge}. Generally, a self-driving vehicle can generate up to 4 terabytes of data per day, and the L5 level of autonomous driving requires computing facility that is capable of executing over 2000 tera operations per second \cite{Survey}. In this regard, it becomes paramount to investigate how to achieve highly accurate and low latency on-board inference results with limited computing resources.
	
	Cloud server with powerful GPUs can help to mitigate computation-intensive inference burden on devices \cite{yang23edge, dai22respire}. However, the strict requirements for end-to-end latency of self-driving vehicles makes the cloud-based offloading strategy unfeasible \cite{AutonumousDriving, DAO}. In particular, considering the high-speed motion characteristic of vehicles, edge nodes (ENs) in the proximity are more suitable for vehicular networks. The vehicles can offload part of analytics tasks when it is within the coverage of ENs, which is referred to as road-side units \cite{Survey, he23edge}. Considering the packet loss due to the instability of network connection in the interest of vehicle (IoV), reducing the amount of data transmitted to ENs should be considered without deteriorating inference accuracy, in order to guarantee driving safety. 
	
	The redundant information in the video frames generated by on-board cameras, not only affects detection accuracy, but also increases the transmission burden. The RoI in high-resolution video frames actually takes up only a small fraction of pixels, which provides the possibility for data compression \cite{Remix}. RoI extraction is a classic problem in computer vision, such as region proposal network in faster R-CNN \cite{FasterRcnn} and selecting search strategy in R-CNN \cite{RCNN}. However, these RoI extraction methods are time-consuming and difficult to perform in real time. As a result, we need to explore a low-cost RoI extraction method for background de-redundancy to reduce the amount of transmitted data. Class activation mapping (CAM) \cite{CAM} has the remarkable ability to localize objects with the help of feature map extracted by convolutional layers, which can be used for RoI extraction. More importantly, it only takes tenths of milliseconds to process one image.
	
	In this paper, we investigate lightweight RoI extraction method for autonomous vehicles, and propose an edge-assisted video analytics system to minimize the end-to-end latency while ensuring high-performance accuracy of vision tasks. RoI boxes extracted from high-resolution frame based on the location ability of  CAM are assigned to different down-sampling rates and offloading to edge aiming to lighten the computing burden on autonomous vehicles. The main contributions of this paper can be summarized as follows.
	\begin{itemize}
		\item{We propose an edge-assisted real-time video analytics system for high-resolution video streams generated by on-board cameras, which significantly improves the inference accuracy for vision-based vehicle perception.}
		\item {We design a lightweight CAM-based RoI extraction method, which extracts RoI boxes from high-resolution frames at low complexity.}
		\item {Aiming to strike a balance between inference accuracy and the cost of data processing, we propose an adaptive algorithm, which selects resolution for each valid RoI box based on network fluctuation and available edge resources.}
	\end{itemize}
	
	The remainder of this paper is organized as follows. Section \ref{sec_Motivation} describes the motivation of our design, followed by system model in Section \ref{sec_Pinpeline}. Section \ref{sec_Detail1} presents the adaptive boxes offloading strategy. Performance evaluation is given in Section \ref{sec_PerformEvaluation}. Finally, Section \ref{sec_conclusion} concludes this paper with future work.

	\section{Motivation}\label{sec_Motivation}	
	\begin{figure}[t]
		\centering
		\begin{minipage}[t]{0.24\textwidth}
			\centering
			\includegraphics[width=\textwidth]{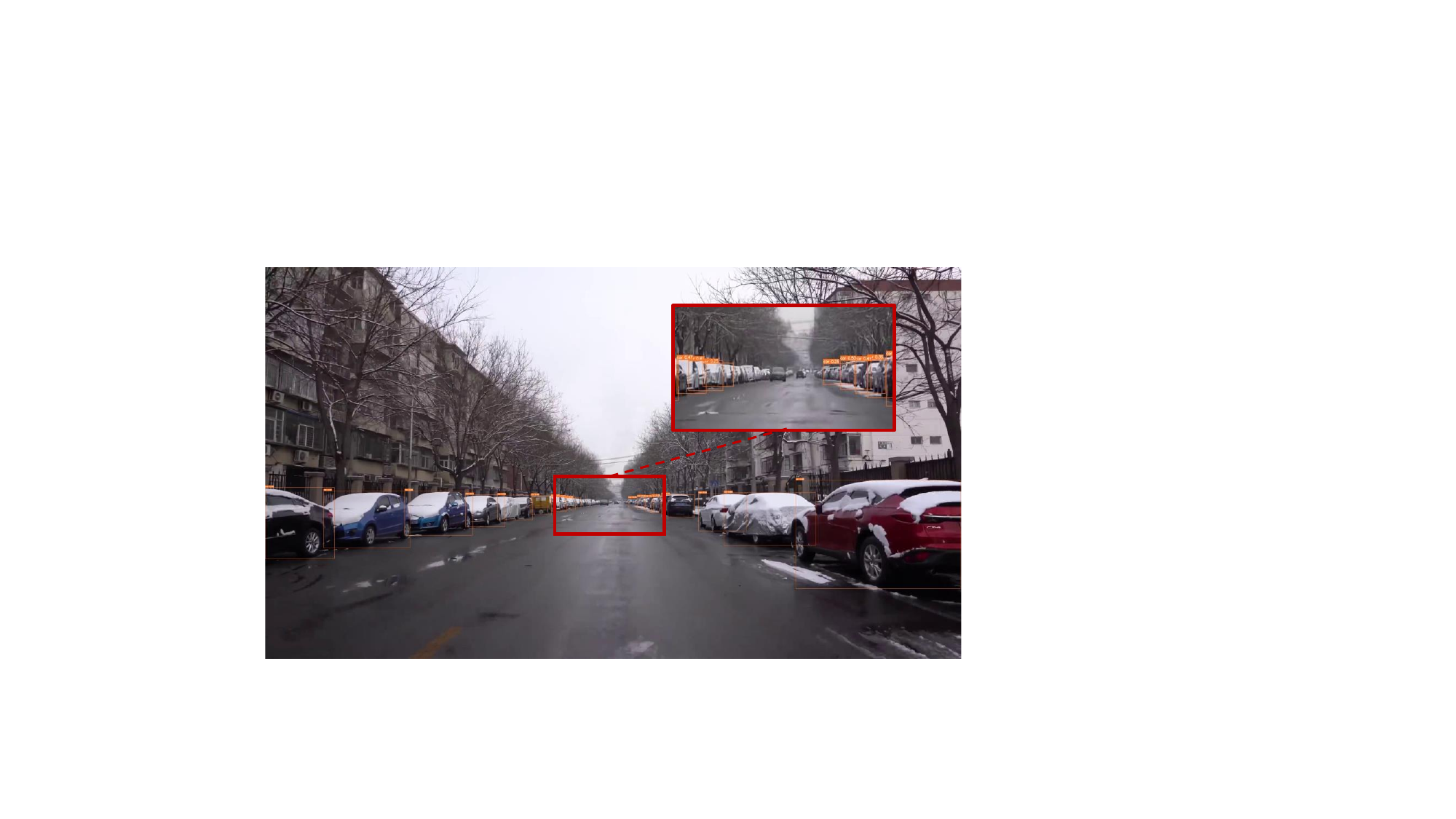}
			\centerline{(a) }
		\end{minipage}
		\hfill
		\begin{minipage}[t]{0.24\textwidth}
			\centering
			\includegraphics[width=\textwidth]{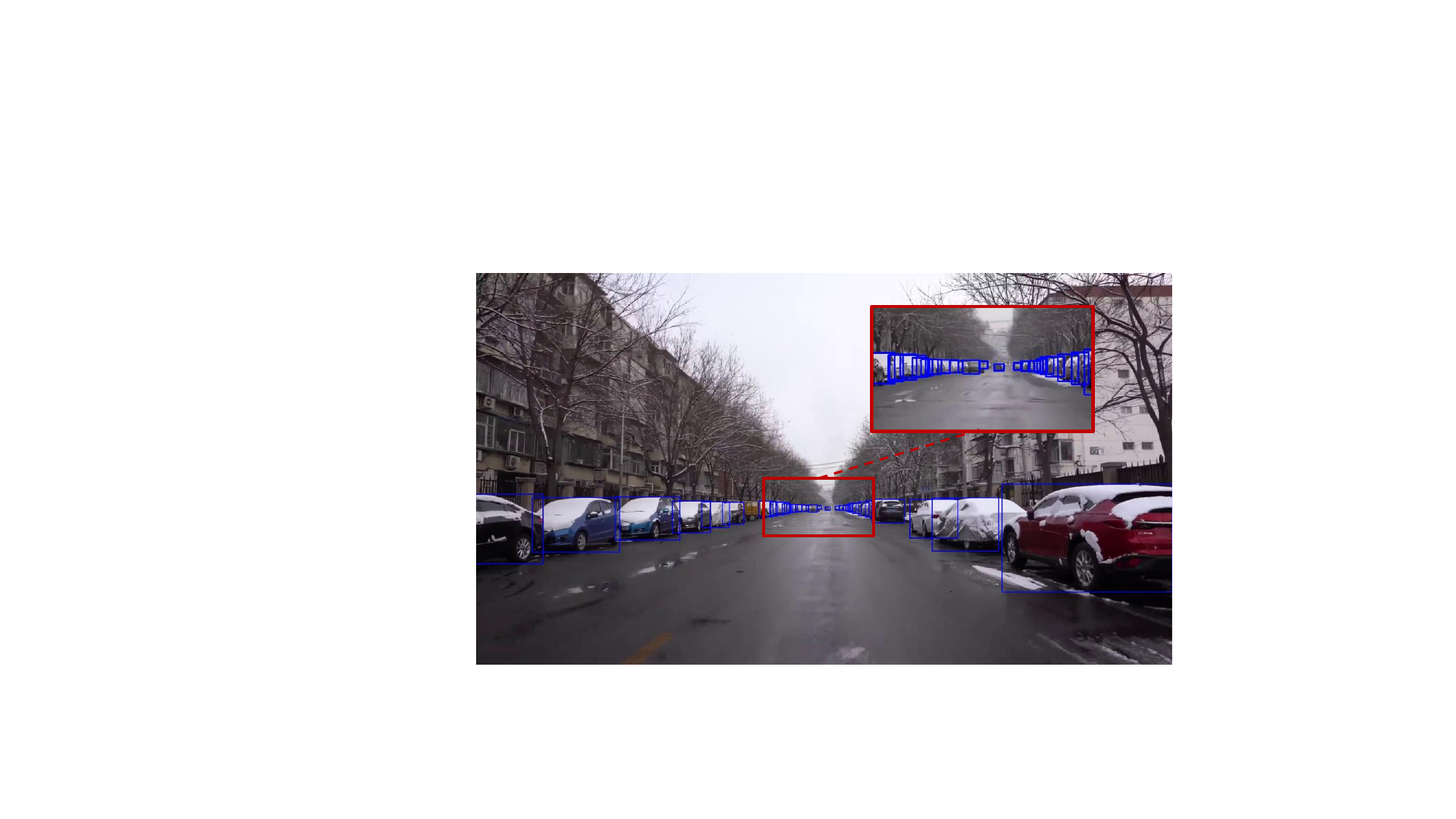}
			\centerline{(b) }
		\end{minipage}
		\caption{Comparison of object detection from high-resolution video, (a) input 4K raw image and (b) input after CAM-based RoI extraction.}   
		\label{motivation} 
		\vspace{-0.4cm}     
	\end{figure}
	
	\subsection{On-Board Processing of High-Resolution Videos}
	High-resolution images and videos are with much more pixels, which provide rich on-road information for safe autonomous driving\cite{AdaMask}. The feature maps extracted by CNNs are semantically richer and the correlation position relationships between objects are spatially more accurate. However, there are few lightweight CNNs designed for real-time processing of high resolution images. The datasets used for training CNNs are mostly small size images, such as 224$\times$224 or 640$\times$640, which may not be applicable to vision tasks on 4K or 8K images \cite{Remix}. High resolution video streams need to be downsampled before being input to CNNs, leading to accuracy drop. The architecture of CNNs could also be scaled up and trained on high-resolution datasets, at the cost of unacceptable latency \cite{Bhardwaj22ekya, kong23edge}. It takes about 49 millisecond (ms) for object detection task of a 4K image using YOLOv5x (1280) \cite{YOLO} on an RTX 3090 GPU, which is far from achieving the real-time inference on video streams.
	
	\subsection{CAM-based Frame Partitioning}
	Equal image partitioning is an effective way to improve the prediction accuracy, but it can not work well with limited computation and bandwidth resources\cite{ELF}. Instead of reducing the amount of data transmitted to the edge, image partitioning even increases the computational burden of prediction by several times (each part of the image requires CNNs for processing), leading to high end-to-end latency. On the other hand, image partitioning without incorporating object distribution information leads to the loss of accuracy for objects at the boundary of the partitions, and the waste of resources in areas where no object exist. Therefore, we consider RoI extraction based on object location algorithm to reduce the large volume of redundant data transmitted, and improve the prediction accuracy without increasing the input size of CNNs.  
	
	In order to demonstrate the advantages of image partition, we conduct experiments on a computer with RTX 3090 GPU, and employ YOLOv5x (640) for object detection task on a 4K YouTube video. As shown in Fig. \ref{motivation}, the detection accuracy of our proposed system is significantly higher compared to directly feeding raw image to the CNNs. Fig. \ref{motivation}(b) shows the result of using CAM-based RoI extraction, which can effectively improve the inference accuracy. Hence, in this paper, we investigate low-complexity CAM-based RoI extraction method to achieve fast and high-accuracy vehicle perception.

	\section{System Model and Problem Formulation}\label{sec_Pinpeline}
	\subsection{System Overview}
	
	\begin{figure}[!t]
		\centerline{\includegraphics[width=\columnwidth]{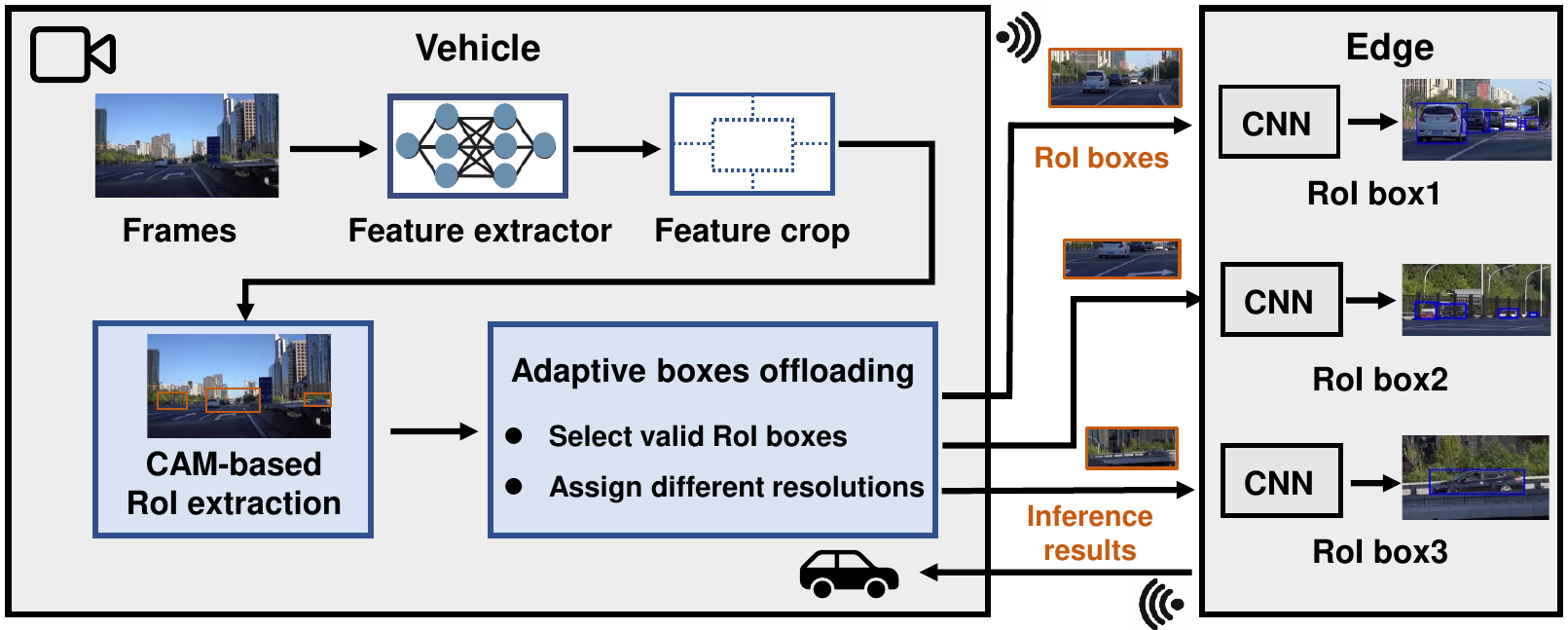}}
		\caption{An overview of the proposed system.}
		\label{pipeline}
		\vspace{-0.4cm}  
	\end{figure}
	
	\begin{figure*}[t]
		\centering  
		\subfigure{   
			\begin{minipage}{5.3cm}
				\includegraphics[scale=0.16]{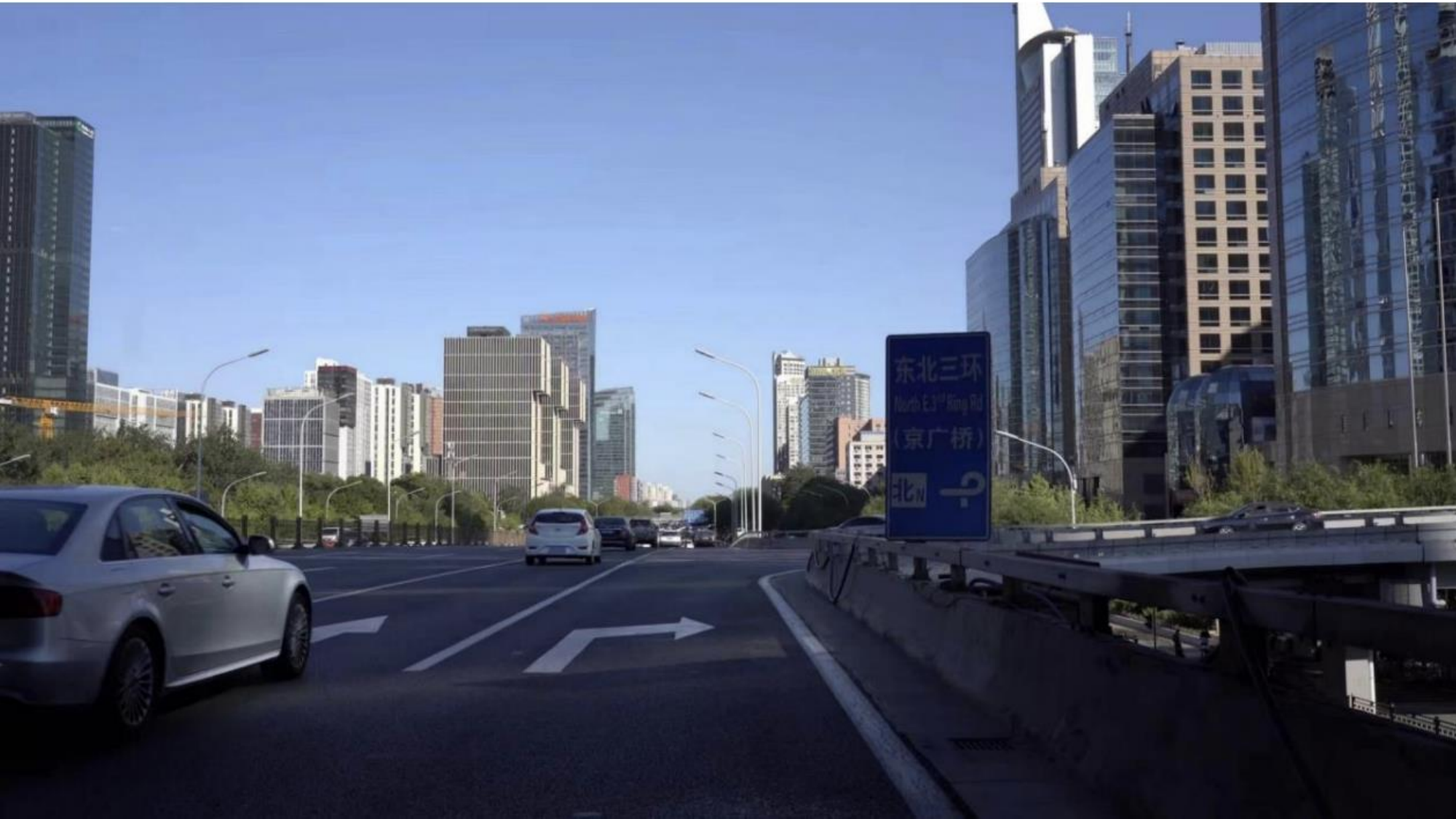}  
				\centerline{(a) Raw image}
			\end{minipage}
		}
		\subfigure{ 
			\begin{minipage}{5.3cm}
				\centering    
				\includegraphics[scale=0.16]{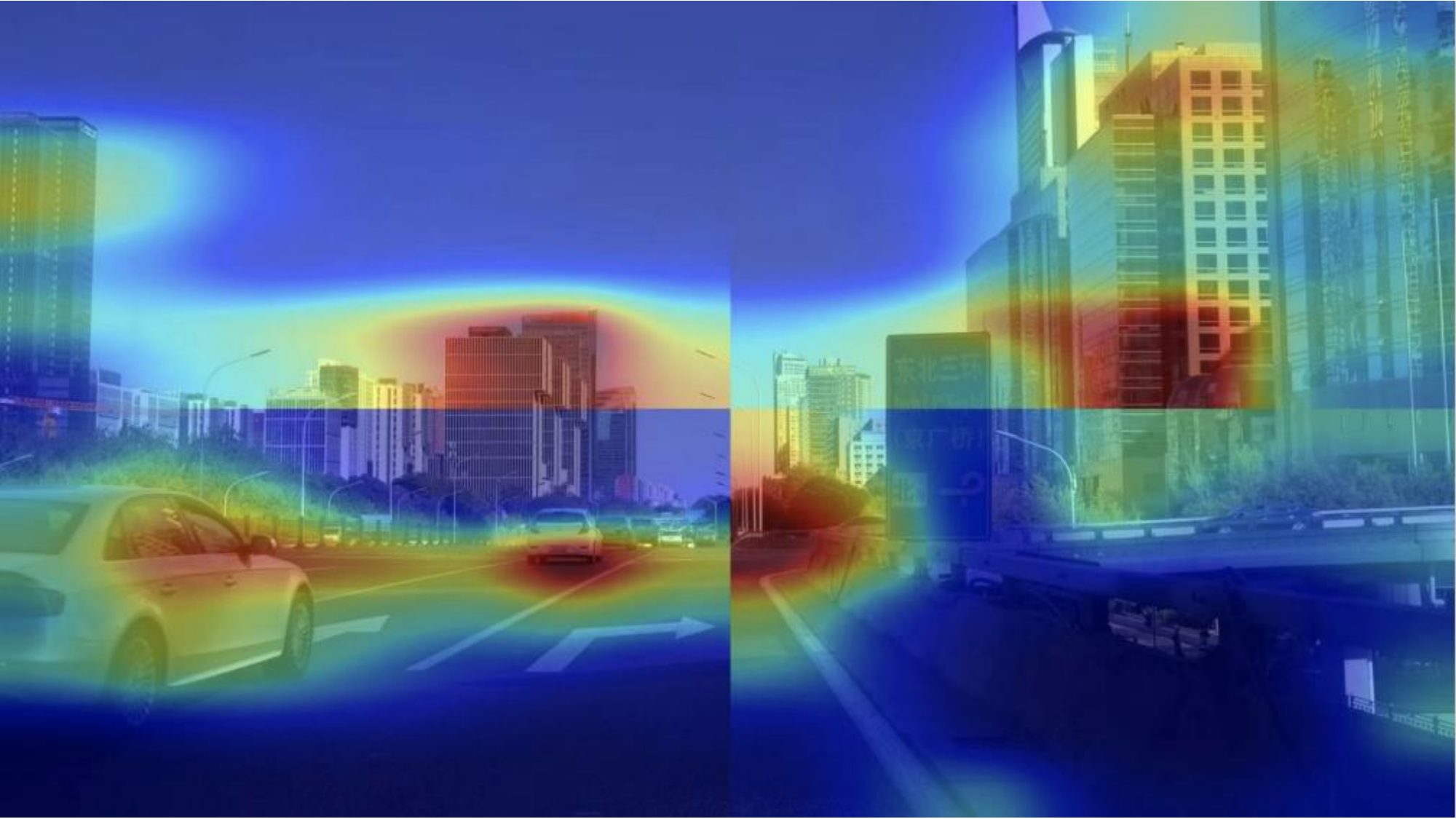}
				\centerline{(b) CAM}
			\end{minipage}
		}
		\subfigure{   
			\begin{minipage}{5.3cm}
				\centering   
				\includegraphics[scale=0.16]{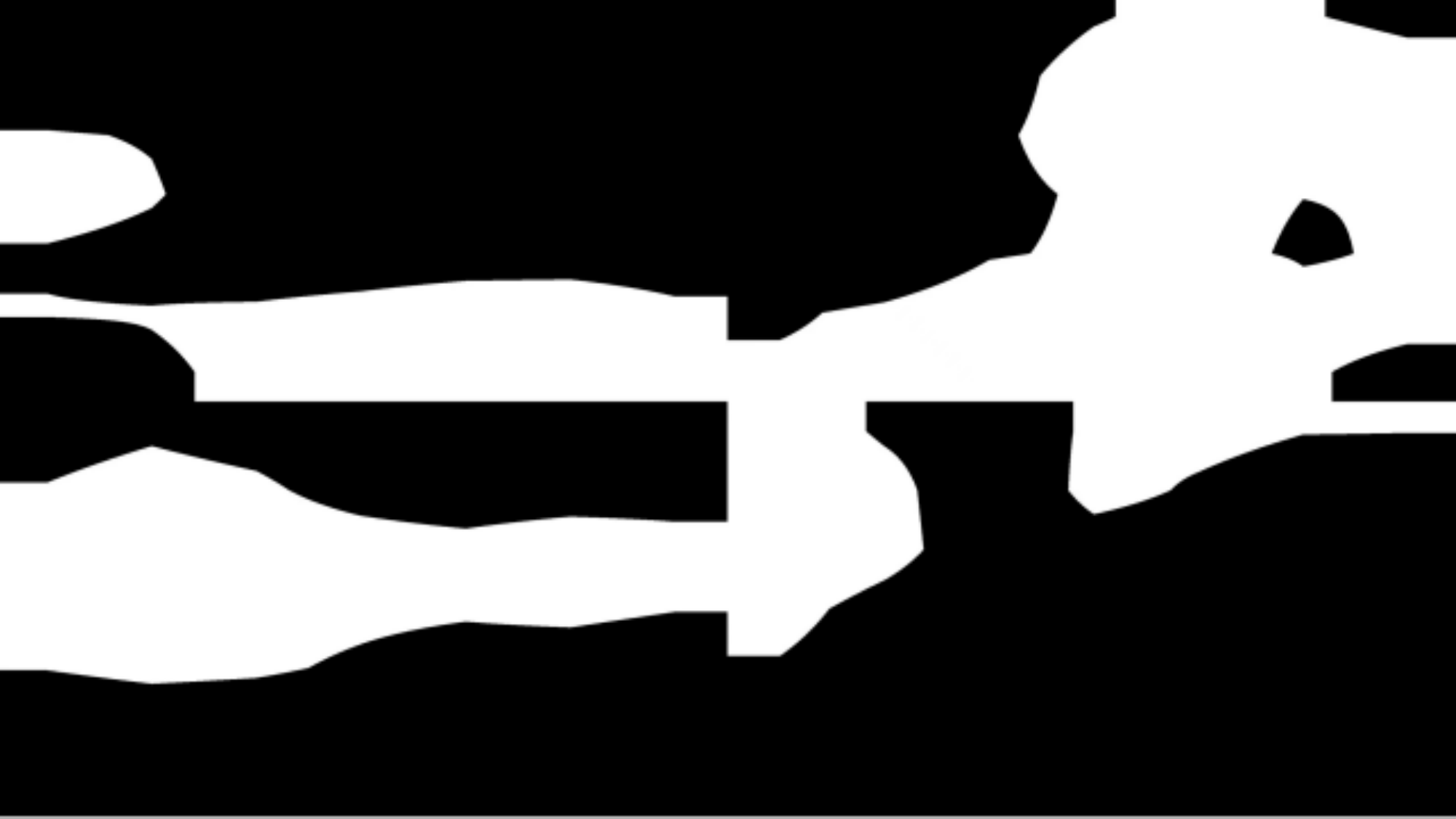}  
				\centerline{(c) Masking}
			\end{minipage}
		}
		\subfigure{ 
			\begin{minipage}{5.3cm}
				\centering    
				\includegraphics[scale=0.16]{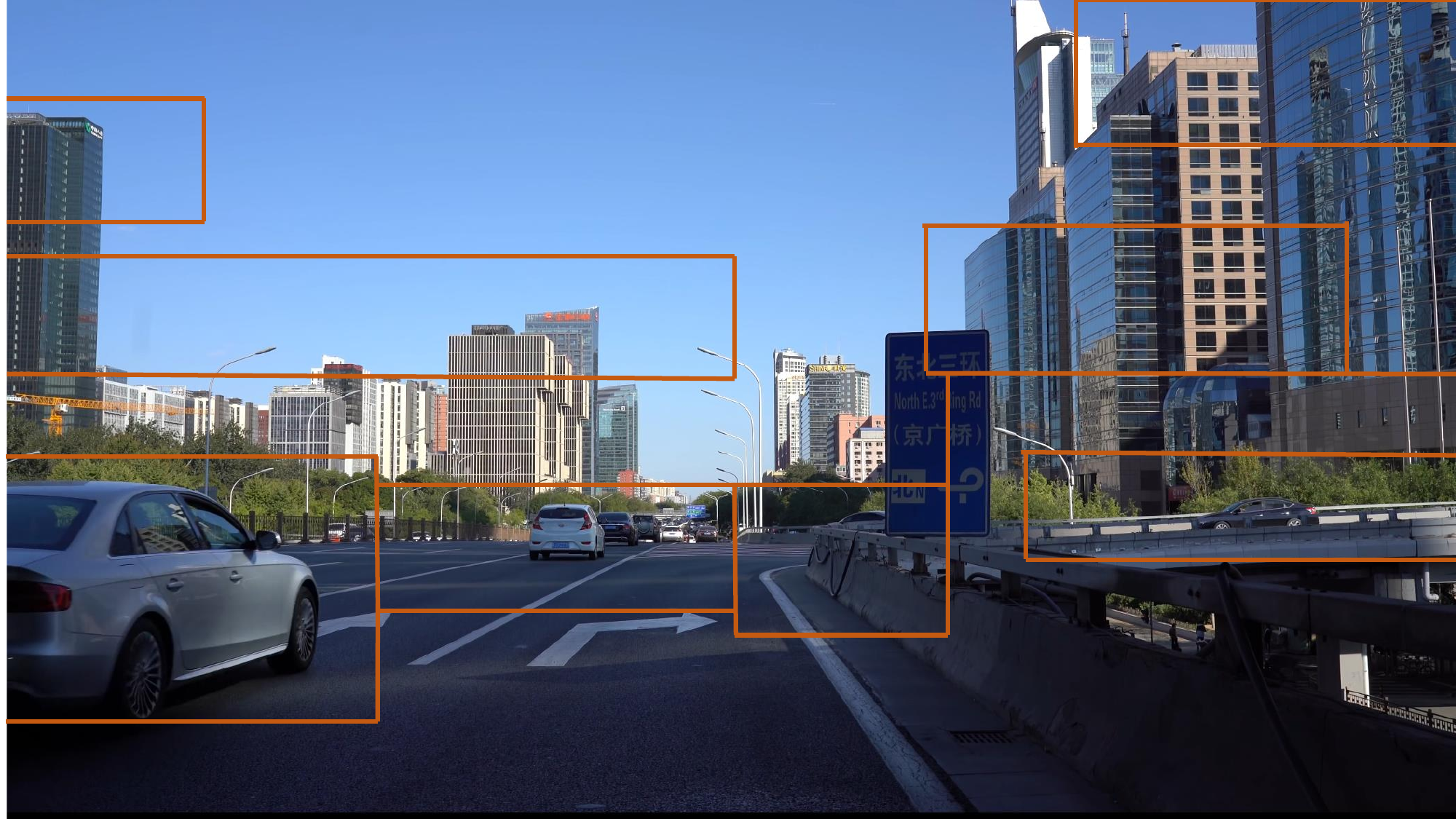}
				\centerline{(d) RoI extraction}
			\end{minipage}
		}
		\subfigure{ 
			\begin{minipage}{5.3cm}
				\centering    
				\includegraphics[scale=0.16]{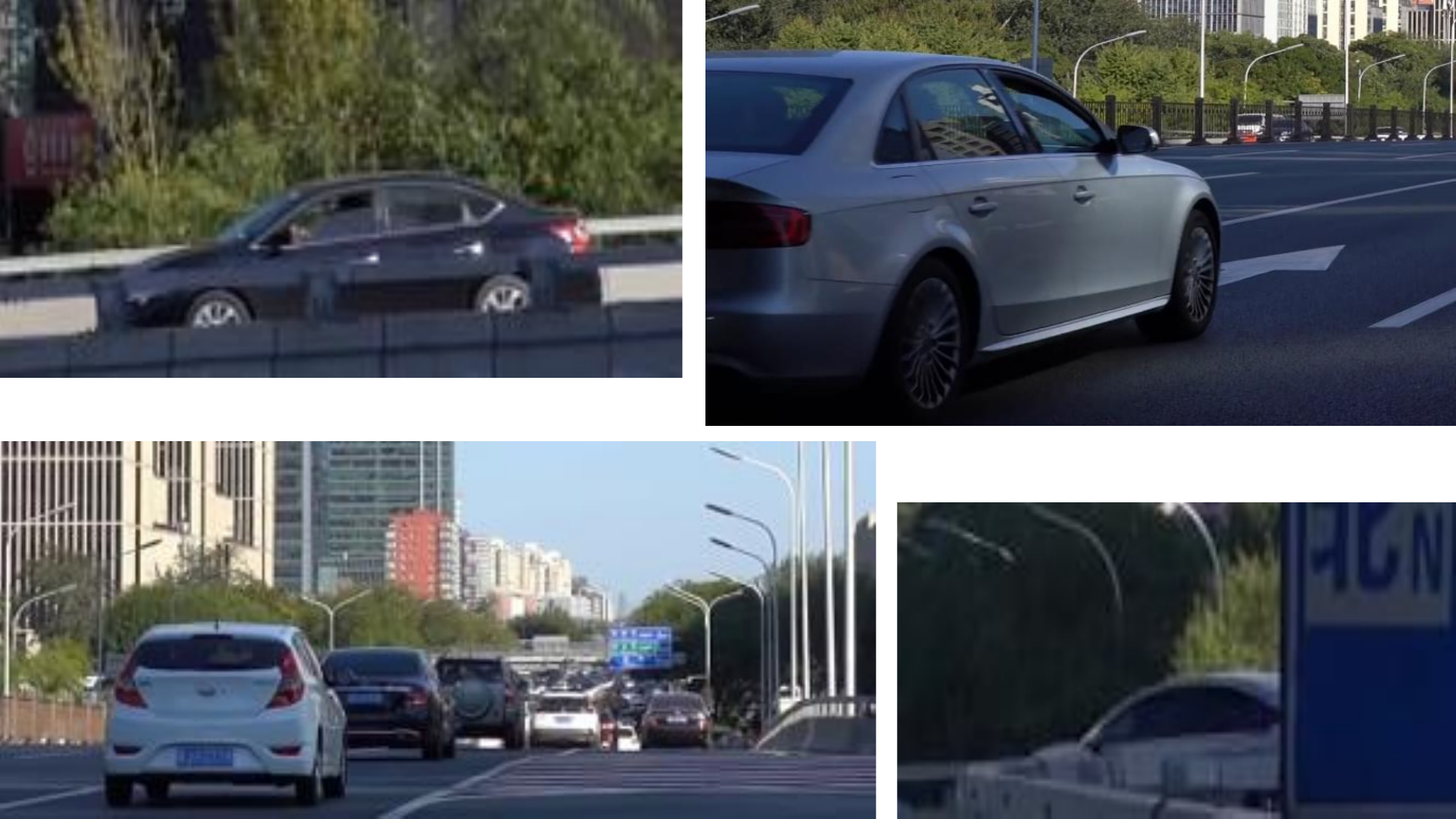}
				\centerline{(e) Resizing of valid RoI boxes}
			\end{minipage}
		}
		\subfigure{ 
			\begin{minipage}{5.3cm}
				\centering    
				\includegraphics[scale=0.16]{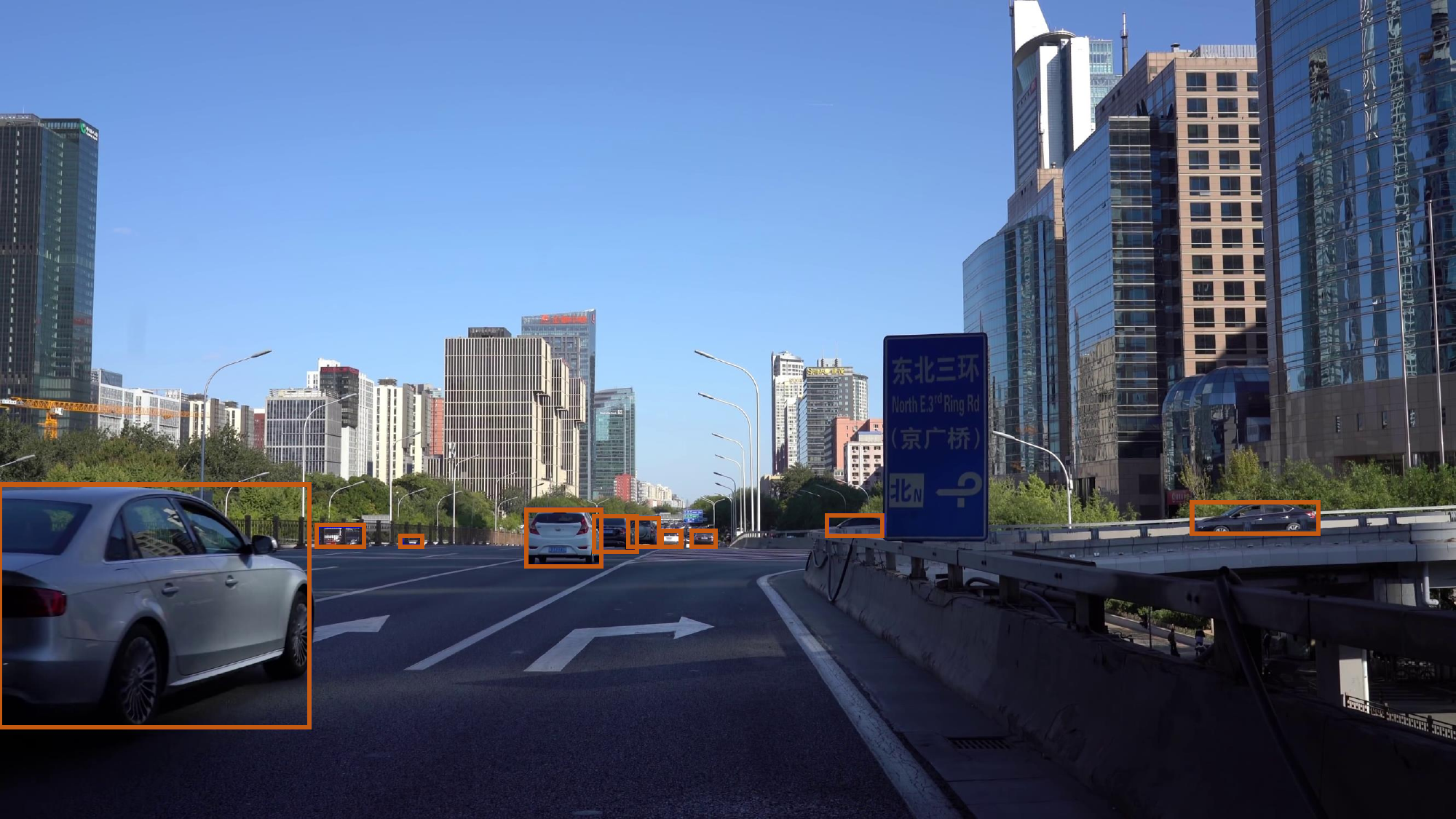}
				\centerline{(f) Final result}
			\end{minipage}
		}
		\caption{(a) Raw image, (b)-(d) The pipeline of CAM-based RoI extraction, (e) an illustration of adaptive resizing of RoI boxes, and (f) result feedback.}
		\label{process}
		\vspace{-0.5cm}  
	\end{figure*}
	
	As shown in Fig. \ref{pipeline}, the edge-assisted real-time video analytics system consists of three components: feature extraction and cropping, CAM-based RoI extraction, and adaptive RoI box offloading. In order to reduce the computing consumption of vehicles and the amount of data transferred to the edge, we employ a lightweight feature extractor and a fixed-mode feature partition method for video frame pre-processing. Then, considering the limited on-board computing capacity, a lightweight CAM-based RoI extractor is utilized on each feature cropping to obtain a set of boxes containing the target objects and regardless of the background region. Before offloaded from vehicles to ENs, the RoI boxes are re-selected to remove invalid boxes and appointed to different down-sampling rates based on the current available bandwidth and edge resources to better balance the inference accuracy and transmission cost. In the rest of this section, we will introduce the specific details of the main component.
	
	\subsection{Feature Extraction and Cropping}
	The CAM has remarkable object location abilities without training on any pixel-level labels \cite{CAM}. For a given image $ \bold{X} \in \mathbb{R}^{D \times (H \times W)}$, let $ F_c(x,y) $ represents the feature of channel $c$ extracted from the last convolutional layer of the lightweight CNN at spatial location $ (x, y) $. After global average pooling layer, we can get a set of scalars $ \{ f_1, f_2, ..., f_n\} $, each of which corresponding to the mean value of a feature map $f_c = \sum_{x,y} {F_c(x,y)} $. Combined with the weight $ \omega_c^i $, we can obtain the class activation map $M_i$ of class $i$: 
	\begin{align}\label{cam}
		M_i  = \sum_i \omega_c^i \cdot f_c,
	\end{align}
	where $ \omega_c^i $ reflects the importance of feature map in corresponding channel $c$ for class $i$. Moreover, we can get the contribution value of each pixel to the class $i$, $M_i(x,y)$. Larger $M_i(x,y)$ indicates the area where the object is more likely to appear, thus enabling the object location based on CAM. However, it is revealed in our experiments that CAM is not good at localizing multiple objects, especially when the objects are small and distributed discretely. The missed target objects will lead to a deterioration in the overall accuracy. Therefore, in order to maximize the performance of CAM low-cost object localization, we crop the feature map extracted by the lightweight CNN into five parts $ \{F_c^1, F_c^2, ..., F_c^5\} $ and perform CAM on each cropping separately, as shown in Fig. \ref{pipeline}(a). We can get the activation value $ M_i^j $ of each part. Compared with employing CAM on the raw image directly, our method conducts feature extraction only once, while the later requires four times more computational cost.
	
	On the one hand, convolution operation does not change the exact spatial mapping relationship between objects. Therefore, convolution followed by cropping the feature map yields the same result, as partitioning the original image first and then extracting the features separately on each part. On the other hand, CAM-based object localization on each feature cropping can help to seize discrete objects scattered in each part of the image.
	In conclusion, combining feature cropping with CAM for object positioning allows for high accuracy of RoI extraction with as less computational resources as possible.
	
	\subsection{CAM-based RoI Extraction}
	After employing CAM on each feature map cropping, we can get the heatmap which shows the localization of target objects (Fig. \ref{process}(b)). Areas with higher heat values $M_i(x,y)$ indicate that the object is more likely to appear. Therefore, by setting a heat value threshold $\sigma_m$, we can obtain the mask $ \bold{A} ^ {\{M_i(x,y) \ge \sigma_m\}} $ to extract the RoI, illustrated as Fig. \ref{process}(c). A larger threshold will lead to smaller extracted regions, which will result in the omission of key information and further reducing the overall accuracy. On the contrary, if we choose a small threshold, the extracted RoI will contain a lot of redundant background information. The small proportion of the target object in the image will lead to an increase in the difficulty of the visual task, and redundant information will also reduce the efficiency of RoI extraction, resulting in an increase in the amount of transmitted data. Additionally, retaining a small portion of the background area helps improve the accuracy of the visual task.
	
	In short, we choose an empirical threshold to balance the amount of information extracted and the efficiency of region extraction. After region segmentation, we can get a set of RoI boxes as Fig. \ref{process}(e), which can not only reduce the amount of data transferred, but also improve inference accuracy.

	\subsection{RoI Boxes Selection and Offloading}
	The upper part of Fig. \ref{process}(b) illustrates the poor localization performance of CAM over the whole background area, the RoI boxes extracted from which are most invalid boxes without any target objects. Offloading invalid boxes to the servers not only wastes bandwidth resources but also increases the computational burden at the edge, thus we adopt a RoI box selection algorithm to remove the invalid boxes that do not contain any target objects. 
	
	According to our experimental results, approximately 2 to 6 valid RoI boxes will be extracted from the raw frame through our RoI box selection algorithm, with each box ranging in size from 20 KB to 500 KB. On the one hand, due to non-visual range and high mobility of vehicles, the IoV is unstable and thus leads to the packet loss and untimely update of inference results. On the other hand, predicting RoI boxes increases the computational burden of the edge by several times compared to inputting the raw video frame directly which put computation pressure on resource-constraint edge servers. Lower resolution of the boxes leads to low prediction latency of the CNNs deployed on the edge at the cost of accuracy deterioration. Therefore, to mitigate the impact of network fluctuations on computation task offloading and maximize the prediction performance of CNNs at the edge, we propose a adaptive resizing strategy, which determines the resolution of each box by adjusting the down-sampling rate according to the current available bandwidth and computation resources at the edge.
	
	Generally, image resolution can largely affect the inference accuracy and profiling cost of video analytics tasks. For example, the higher resolution of each box, the more accurate inference results obtained by the CNNs, but it leads to longer prediction latency and more GPU utilization. To measure the performance of CNNs on processing the RoI boxes extracted from a video frame, we define a utility function:
	\begin{align}
		u^t = \sum_{i = 1}^{N} a_{r_i}^t- \omega \sum_{i = 1}^{N} c_{r_i}^t, \forall t \in \mathcal{T},
	\end{align}
	where $N$ denotes the total number of valid RoI boxes at time $t$. $a_{r_i}^t$ and $c_{r_i}^t$ represent the accuracy and resource consumption in predicting the $i$-th box with down-sampling rate $r_i$ respectively. The resource consumption consists of two components, i.e., inference latency and GPU utilization of the edge server. 
	
	\subsection{Problem Formulation} 
	Our objective is to maximize the utility function under network bandwidth and edge-side resources constraints:
	\begin{align}
		\mathcal{P} : &  \max_{r_i} u^t,\\
		\emph{s.t.} \,\, & C_1: \sum_{i = 1}^{N}S_{r_i} ^t \leq B^t,\,\,\forall t\in\mathcal{T},\\
		&  C_2: \sum_{i = 1}^{N}G_i ^t \leq G_{\max} ^t,\,\,\forall t\in\mathcal{T}, \\
		&  C_3: r_i \geq r_{\min}. 
	\end{align}
	where $B^t$ and $G_{\max} ^t$ denote the current maximum available bandwidth and GPU resources, respectively. Constrain $C_1$ ensures that the total data volume transmitted per frame does not exceed the current available bandwidth. And constrain $C_2$ represents that the total GPU usage required for the RoI boxes inference is no more than the total available GPU resources at the edge. When the box resolution is smaller than the input size of CNN model, the inference accuracy will be significantly reduced, so we set a lower bound of resolution $r_{\min}$ and constrain $C_3$. We discretize the variables by restricting the value of the downsampling rate $r_i$ and adopt hill climbing algorithm to solve the optimization problem $\mathcal{P}$.

	\begin{figure}[t]
		\centering  
		\subfigure{   
			\begin{minipage}{0.145\textwidth}
				\centering   
				\includegraphics[width=\textwidth]{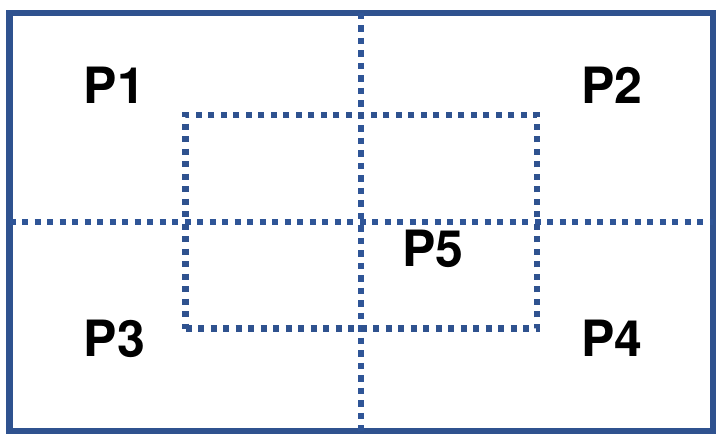}  
				\centerline{(a)}
			\end{minipage}
		}
		\subfigure{ 
			\begin{minipage}{0.145\textwidth}
				\centering    
				\includegraphics[width=\textwidth]{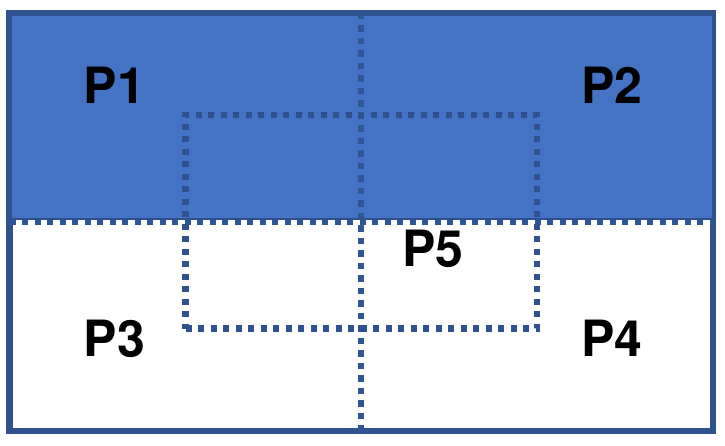}
				\centerline{(b) }
			\end{minipage}
		}
		\subfigure{ 
			\begin{minipage}{0.145\textwidth}
				\centering    
				\includegraphics[width=\textwidth]{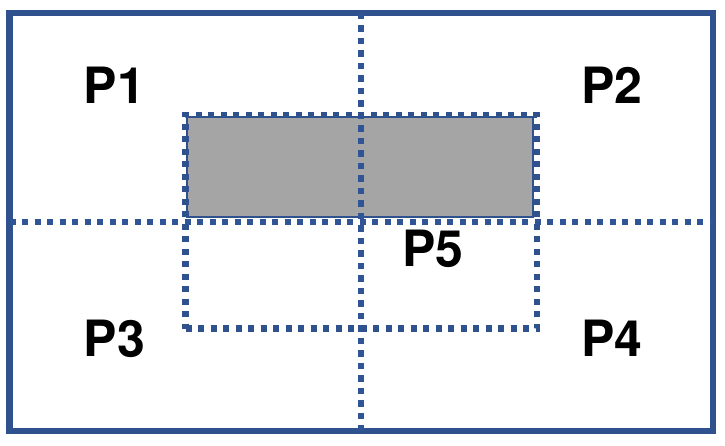}
				\centerline{(c)}
			\end{minipage}
		}
		\caption{(a) An illustration of ``4+1'' partitioning. (b) Calculating the probability of objects appear in $P1 \cup P2$ (blue area). (c) Calculating the probability of objects appear in $P5 \cap (P1 \cup P2) $ (grey area).}   
		\label{partition} 
		\vspace{-0.3cm}
	\end{figure}
	
	\begin{table}
		\caption{The probability of object occurrence in each part}
		\label{table}
		\centering
		\setlength{\tabcolsep}{4pt}
		\begin{tabular}{c|c|c|c|c}
			\hline
			Parts & Dataset1 & Dataset2  &  Dataset3 & Dataset4 \\
			\hline
			\hline
			$P1 \cup P2 $ & 1.008$ \%$  & 1.146$ \%$  & 19.531$ \%$ & 1.692$ \%$ \\
			\hline
			$P5 \cap (P1 \cup P2) $ & 0.129$ \%$  &  1.076$ \%$ & 7.513$ \%$  & 0.866$ \%$ \\
			\hline
		\end{tabular}
		\label{probability}
		\vspace{-0.4cm}  
	\end{table}
	
	\section{Adaptive Boxes Offloading Strategy}\label{sec_Detail1}
	In order to maximize the performance of CAM-based RoI extraction, we crop the feature map. According to the picture structure information of the video captured by the on-board cameras, it can be seen that the middle area of the video frames has the most important information and the densely distribution of objects. Hence, we adopt a ``four plus one'' cropping method shown as Fig. \ref{partition}(a). Note that the partitions can be in non-equal proportions, only the ``four plus one'' mode is fixed. The proportion of each part will be determined offline according to the focal length of the on-board cameras. Cameras with large focal length result in a small view range of video frames. Thus we can reduce the proportion of Part1 (Part refer to as P for simplicity in the rest of the paper) and P2. In the following experiments, we adopt the method of dividing in equal proportions which can cover most scenarios, namely, each part is in the same size.
	
	The P1 and P2 are mainly background areas such as sky and buildings, RoI boxes extracted from them are always invalid. We count the probability of the target object appearing in P1 and P2 (as shown in Fig. \ref{partition}(b)), as well as the part of their concatenation with P4 (as shown in Fig. \ref{partition}(c)) under four different in-vehicle 4K video datasets derived from YouTube. As we can obtain from Table \ref{probability}, in addition to the high probability of object appearing in the upper half of the video frame under dataset3, the probability of missing useful information caused by discarding P1 and P2 can be controlled within 1$ \% $ in the rest datasets. The high probability of object appearing in dataset3 can be solved by expending the proportion of P3 and P4. When we adjust the ratio of the top half to the bottom half to 1:3, the probability of detecting the target object in P1 and P2 of dataset3 will be less than 0.1\%. As a result, we do not need to perform CAM on P1 and P2.
	
	\begin{algorithm}[t!]
		\caption{Adaptive boxes offloading algorithm}
		\label{Alg1}
		\LinesNumbered
		\KwData{RoI boxes sets $\{\mathbf{b}_k\}, k \in \{3, 4, 5\}$;}
		\KwResult{Select valid RoI boxes and assign the optimal resolution to each box;}
		\tcc{Initialize with the highest transmission frequency} 
		$fre_k \leftarrow initFre$; \\
		\tcc{Adjust the transmission frequency}
		\While {$b_i \in \mathbf{b}_k$} {	
			$r_i \leftarrow getResolution(b_i)$; \\
			$res \leftarrow collectionResult(\mathbf{b}_k)$; \\
			\If{$res = 0$ {\bf and} $fre_k \neq 0$} {
				$fre_k \leftarrow fre_k - freIntvl$; \\
				\Else {
					$fre_k \leftarrow initFre $; \\
				}
			}
		} 
		\vspace{-0.1cm} 
	\end{algorithm} 
	
	As for the invalid boxes extracted from P3, P4 and P5, we introduce an algorithm to control the transmission frequency of each part, with the aim of further reducing the amount of data transmitted and the computation burden of the edge, which is shown in Algorithm \ref{Alg1}. Specifically, all extracted RoI boxes are initially offloaded to the edge with the highest frequency of 30 frames per second (FPS). If there is no object detected in a part, the offloading frequency of this part is reduced by five frames. And we set the lower bound of the frame transmission frequency to 1 FPS (the video frame rate is 30 FPS). Moreover, once objects are detected in the RoI boxes offloaded to the edge, the transmission frequency will be reset to 30 FPS.

	\section{Performance Evaluation}\label{sec_PerformEvaluation}
	We evaluate the proposed CAM-based RoI extraction and offloading system on multiple video datasets captured by in-vehicle cameras. Key takeaways are as follows:
	\begin{itemize}
		\item[$\bullet$] The proposed system can improve the overall accuracy of object detection task by up to 16\% and data compression rate by 49\% compared with other baselines.
		\item[$\bullet$] The system can adapt well to bandwidth changes, it can improve the inference accuracy by up to 23\% in poor network condition.
	\end{itemize}
	
	\subsection{Experimental Setting}
	\textbf{Platforms.} We implement the CAM-based RoI extraction and adaptive boxes offloading on the NVIDIA Jetson Tx2 (256 CUDA cores) which is considered to be the on-board processor with considerable computation resources. And the edge server runs Ubuntu and has one NVIDIA GeForce RTX 3090 GPU (10496 CUDA cores) and Intel Core i9-10900K CPU. 
	
	\textbf{CNN Models and Datasets.} To validate the ability of our RoI extraction system in handling computer vision tasks, we choose a common autonomous driving task and employ state-of-the-art CNN models for inference, i.e., the YOLOv5x (640) will be used for object detection on the edge server. And the lightweight feature extractor is chosen as ResNet18. The test datasets are derived from YouTube containing four different video streams generated by in-vehicle cameras. We use the original encoded MP4 format (3840$\times$2160 resolution, 30fps frame rate) as the input.
	
	\textbf{Baseline.} We compare our proposed system with the following video analytics pipelines.
	\begin{itemize}
		\item[$\bullet$] $\mathsf{ELF}$ \cite{ELF} offloads the low-resolution video frame every three frames to edge servers to handle new-occurred objects. The region proposals can be obtained via an attention-based Long Short-Term Memory (LSTM) prediction network and the previous inference results.
		\item[$\bullet$] $\mathsf{EdgeDuet}$ \cite{EdgeDuet} partitions the image into tiles equally and offloads the tiles containing small objects to the remote CNN models located on the edge servers, with the aim of improving detection accuracy.
		\item[$\bullet$] $\mathsf{LEB}$ is a variant of our proposed system, which only performs \textbf{l}ocal CAM-based RoI \textbf{e}xtraction and \textbf{b}oxes selection. The RoI boxes will be fed into the local CNN model for inference. Moreover, in order to ensure real-time inference on the local side, we choose the YOLOv5s (640) model for object detection task to run on TX2.
		\item[$\bullet$] $\mathsf{Origin}$ directly inputs the raw video frames into the CNN model without image partition. 
	\end{itemize}	 
	
	\subsection{Performance Comparison}
	In this section, we evaluate our proposed video analytics pipeline under four different datasets and varying bandwidth conditions, and compare it with other baselines. We employ the F1 score \cite{wang22object} to measure the object detection performance and YOLOv5x (1280) placed on the server to obtain the ground truth. The performance is evaluated by the mean accuracy and the frame size after processing with different algorithms.
	
	\begin{figure}[t]
		\centering
		\begin{minipage}[t]{0.24\textwidth}
			\centering
			\includegraphics[width=\textwidth]{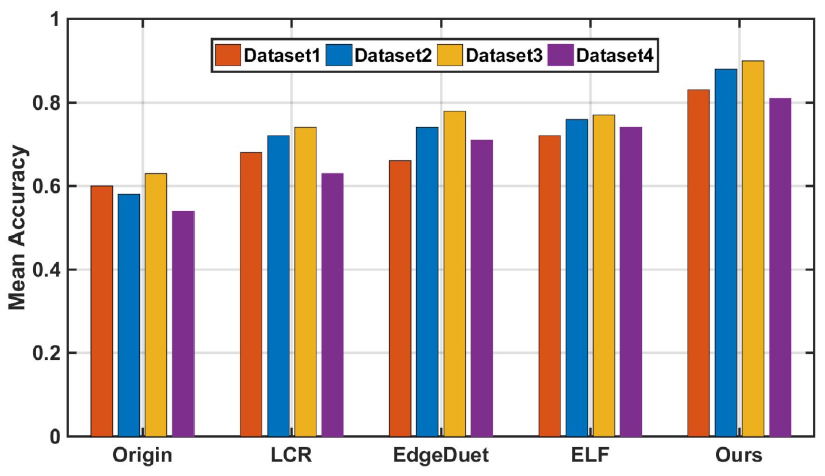}
			\centerline{(a)}
		\end{minipage}
		\hfill
		\begin{minipage}[t]{0.24\textwidth}
			\centering
			\includegraphics[width=\textwidth]{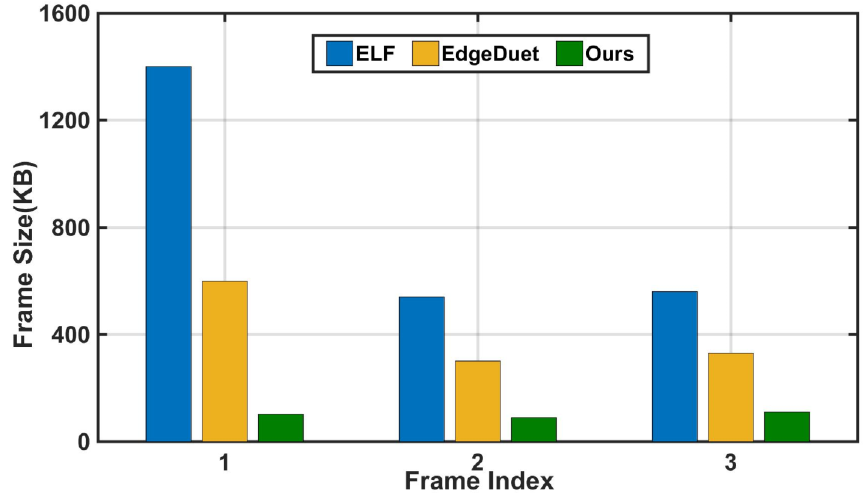}
			\centerline{(b) }
		\end{minipage}
		\caption{Comparison of different methods in (a) Mean inference accuracy and (b) data compression ratio.}   
		\label{accuracy}    
		\vspace{-0.4cm} 
	\end{figure}

	The results are shown in the Fig. \ref{accuracy}, where Fig. \ref{accuracy}(a) is the comparison of mean accuracy between our proposed system and other baselines under four different on-board datasets. Compared to other algorithms, our proposed system obtains the highest average accuracy under all datasets. Although both $\mathsf{ELF}$ and $\mathsf{EdgeDuet}$ partition the raw image into tiles and handle each tile separately to improve the detection accuracy. However, $\mathsf{ELF}$ obtains region proposals through a prediction network according to the previous inference results of the low-resolution video frame transmitted to the edge at the beginning, which limits the accuracy performance improvement of the whole system. As for $\mathsf{EdgeDuet}$, the equal partitioning scheme can not eliminate background areas very well which further influences the detection accuracy. To make a great improvement in the inference accuracy, our proposed system performs low-cost RoI extraction based on CAM. The result in Fig. \ref{accuracy}(a) shows that compared with $\mathsf{ELF}$, our system achieves up to 16\% mean accuracy increase.
	
	Figure \ref{accuracy}(b) shows the frame size of three consecutive frames, namely, the amount of data transferred under different algorithms. Likewise, our proposed system demonstrates significant advantages in data compression compared with baselines. The frame size after boxes resizing of our proposed system is only about 100 KB (including four valid boxes) of one frame. The data size of the first frame in $\mathsf{ELF}$ is relatively high because the low-resolution frames are transmitted to the edge for processing together with the region proposals. And the data size of $\mathsf{EdgeDuet}$ is related to the frame position in a group of pictures, which is also larger than ours.
	
\begin{figure}[t]
		\centering
		\begin{minipage}[t]{0.23\textwidth}
			\centering
			\includegraphics[width=\textwidth]{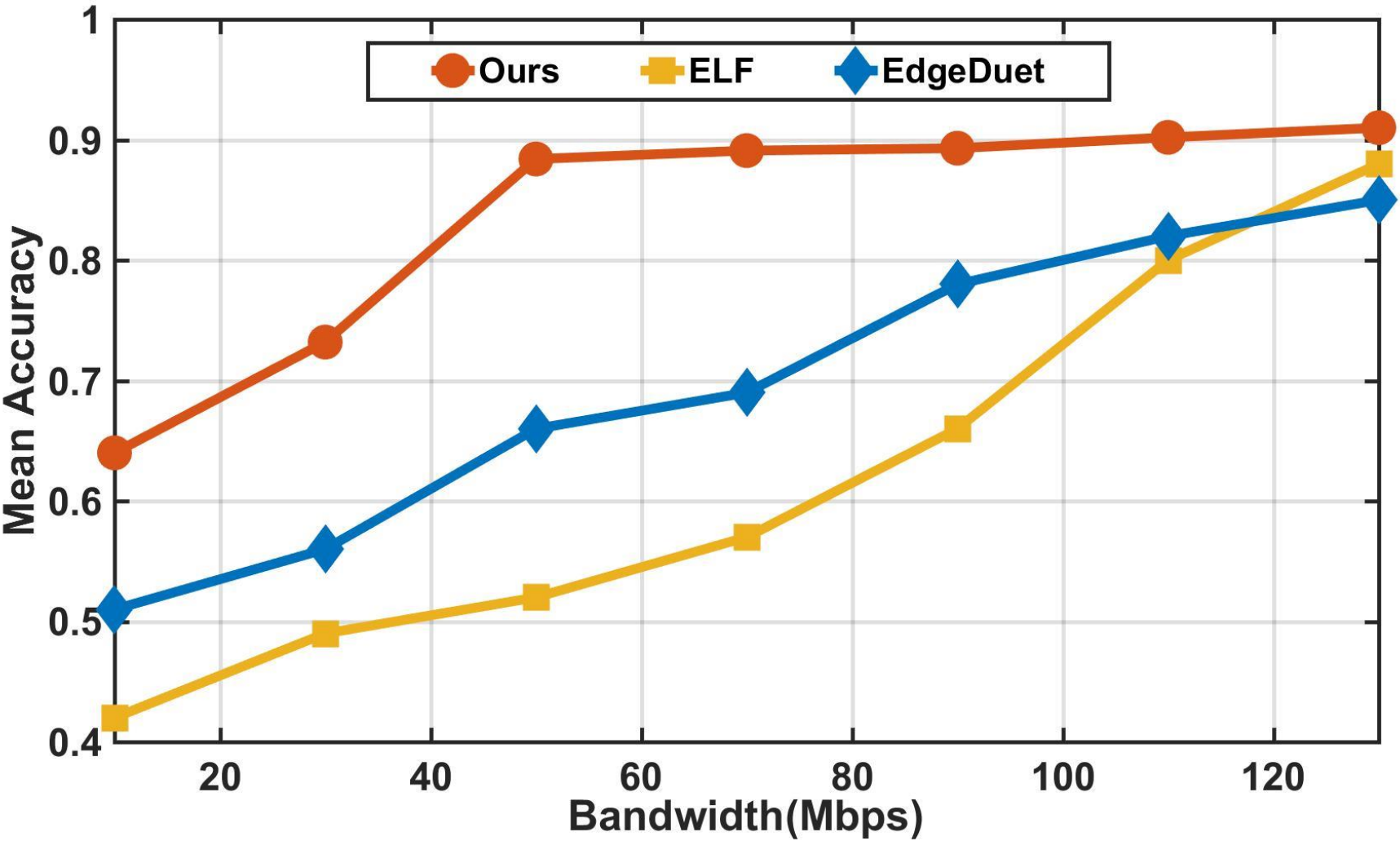}
			\centerline{(a)}
		\end{minipage}
		\hfill
		\begin{minipage}[t]{0.24\textwidth}
			\centering
			\includegraphics[width=\textwidth]{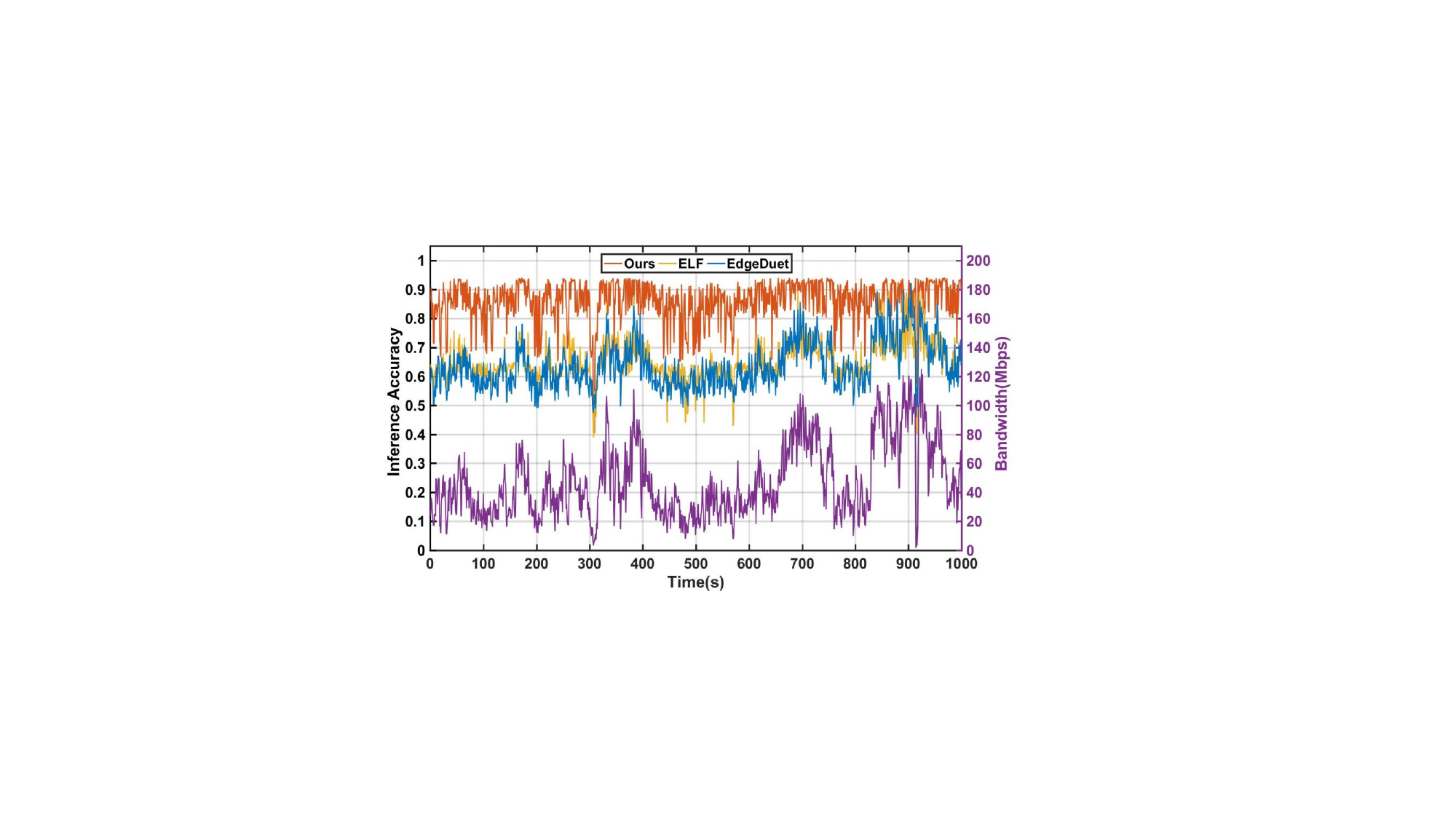}
			\centerline{(b) }
		\end{minipage}
		\caption{An illustration of impact of (a) network bandwidth and (b) bandwidth fluctuation on different methods.}
		\label{bandwidth}    
		\vspace{-0.3cm} 
	\end{figure}

	Vehicles' high mobility lead to intermittent IoV connections, and subsequently packet loss. This calls for adaptability to network conditions in system design. Our model shows great advantages in terms of the adaptability to bandwidth. Fig. \ref{bandwidth}(a) demonstrates that, in the case of poor network conditions, e.g., 20 Mbps, our method can guarantee a mean accuracy of 0.7, while that of $\mathsf{ELF}$ is only 0.45. In Fig. \ref{bandwidth}(b), when network bandwidth fluctuates, the inference accuracy of other baselines change drastically, while our algorithm ensures stable and high inference accuracy. Thanks to the extremely high data compression rate of RoI extraction process, the proposed algorithm can maintain high accuracy under various network conditions.  
	
	Note that, the overall computational overhead of our system consists of 2.94 ms feature extraction, and 2 ms box size selection at the device side. For the edge, different analytics tasks correspond to different processing latency, e.g., target detection takes only 10 ms. Moreover, the size of effective boxes after resize is only about 100 KB, and the time for parallel offloading to the edge is negligible. In summary, our system completion can run in real time.
	
	\section{Conclusion}\label{sec_conclusion}
	In this paper, we have proposed a CAM-based RoI box extraction and adaptive transmission system for vehicle perception, aiming at achieving high accuracy at low edge resource consumption. The system performs feature extraction and cropping on the video frames, and conducts CAM on each cropped images to help extract RoI boxes. Valid RoI boxes are then selected and offloaded to edge servers after box resizing for better inference. Extensive experimental results have demonstrated the advantages of the proposed system. For the future work, we will consider scaling up the system to multi-vehicle collaboration.


	\end{document}